\documentclass{aa}
 
\usepackage{graphicx}
\usepackage{txfonts}
\usepackage{epsfig}

\begin{document}
\title{Discovery of new Milky Way star cluster candidates in the 
2\,MASS point source catalog V. Follow-up observations of the young 
stellar cluster candidates RCW\,87,  [BDSB2003]\,164 and  [DBSB2003]\,172.
\thanks {Based on observations collected with the 6.5m Magellan Baade 
telescope, Las Campanas Observatory and NTT, ESO, La Silla.}
}
\subtitle{}

\author{J.~Borissova\inst{1}
\and
V.~D.~Ivanov\inst{2}
\and 
D.~Minniti\inst{3}
\and
D.~Geisler\inst{4}
}

\offprints{Borissova}

\institute{Departamento de F\'isica y Meteorolog\'ia, Facultad 
       de Ciencias, Universidad de Valpara\'{\i}so, 
       Ave. Gran Breta\~na 644, Playa Ancha, Casilla 53,
             Valpara\'iso, Chile, 
\email{jborisso@eso.org}   
\and        
European Southern Observatory, Ave.\ Alonso de Cordoba 3107, 
        Casilla 19, Santiago 19001,                    
        Chile,  
\email{vivanov@eso.org}               
\and   
Pontificia Universidad Cat\'{o}lica de Chile, Facultad de F\'{\i}sica, 
Departamento de Astronom\'{\i}a y Astrof\'{\i}sica,
Av. Vicu\~{n}a Mackenna 4860, 782-0436 Macul, Santiago, Chile, 
\email{dante@astro.puc.cl}
\and
        Grupo de Astronom\'{\i}a, Departamento de F\'{\i}sica, 
        Universidad de Concepci\'{o}n, Casilla 160-C, Concepci\'{o}n, Chile, 
        \email{doug@kukita.cfm.udec.cl}}
 

\date{Received .. ... 2005; accepted .. ... 2005}

\authorrunning{Borissova et al.}
\titlerunning{Obscured stellar cluster candidates}

\abstract
{Massive clusters are more often found in actively star forming 
galaxies than in quiescent ones, similar to the Milky Way.}
{We have carried out an extensive survey of obscured Milky Way 
clusters to determine whether our Galaxy is still forming 
such objects.} 
{Near-infrared spectral classification of some probable cluster 
members was used to derive the distances to the cluster 
candidates. Isochrone analysis of deep near-infrared 
color-magnitude diagrams allowed us to obtain age and mass 
estimates.} 
{We report the physical parameters of three cluster candidates: 

RCW\,87 is $\sim$25\,Myr old, located at a distance of 
D$\sim$7.6\,Kpc in the general direction of the Galactic 
center. Adding the mass of the suspected cluster members we 
obtain an estimated total cluster mass of $\sim$10300$M_{\odot}$. 
The mid-infrared photometry indicates that some next-generation 
star formation is on-going in the associated H\,{\sc II} region, 
probably triggered by supernovae or stellar wind from the older 
stars in RCW\,87.

The brightest member of [BDSB2003]\,164 is an O5\,V type star, 
based on our spectroscopy. This limits the cluster age to less 
than a few million years. The estimated total mass is $\sim$1760 
$M_{\odot}$ and the distance is D$\sim$3.2\,Kpc. 

[DBSB2003]\,172 lacks central concentration and most likely this 
is not a cluster but an extended star forming region.} 
{The high mass of RCW\,87 -- if confirmed -- puts this object 
in line with Arches and Quintuplet, among the most young massive 
clusters in the Galaxy. Further study is necessary to confirm
this important result.}


\maketitle

\section{Introduction}

Recent all-sky near-infrared (IR) surveys (2\,MASS, Skrutskie et al. 
1997; DENIS, Epchtein et al. 1997; GLIMPSE, Benjamin et al. 2003) 
made it possible to carry out uniform census of highly obscured 
Milky Way clusters. However, as a rule, the determination of the 
physical parameters of these objects requires better-quality data 
than the ones the surveys can offer. This motivated us to carry 
out follow up observations to derive the cluster parameters. Our 
target list was selected from the 2\,MASS Point Source Catalog 
(Ivanov et al. 2002, Borissova et al. 2003) and from the cluster 
candidate catalogues of Bica et al. (2003a,b) and Dutra et al. (2003). 
The detailed studies of about two dozen objects are reported in 
Borissova et al. (2005) and Ivanov et al. (2005). They all are 
young (typically 7-10\,Myr) and significantly smaller 
($\leq$5000\,M$_\odot$) than the Galactic globular clusters and 
than the most massive young clusters known (Arches \& Quintuplet), 
indicating that the formation of massive young clusters is not a
common event in the present-day Milky Way. 

In this paper we report high angular resolution deep near-IR 
imaging and spectroscopy of three more star cluster candidates. 
The next section describes the data and the third section 
discusses the confirmed clusters in detail. The last section is a 
summary of the results.

\section{Observations and data reduction}

The imaging observations were carried out with the PANIC 
(Persson's Auxiliary Nasmyth Infrared Camera) near-IR imager on 
the 6.5-meter Baade telescope at the Las Campanas Observatory. 
The instrument uses a 1024$\times$1024 HgCdTe Hawaii detector 
array. The scale is 0.125 arcsec $\rm pixel^{-1}$, giving a 
total field of view of 2.1$\times$2.1 arcmin. The observing log 
is given in Table~\ref{TableLog}. The observing strategy is 
described in Borissova et al. (2005). 

\begin{table*}
\begin{center}
\caption{Parameters of the cluster candidates and the log of 
observations.}
\label{TableLog}
\begin{tabular}{lccrcrrc}
\hline
\multicolumn{1}{c}{ID} &
\multicolumn{1}{c}{R.A.} &
\multicolumn{1}{c}{Dec.} {\hspace{5pt}}&
\multicolumn{1}{c}{{\it l}}{\hspace{5pt}} &
\multicolumn{1}{c}{{\it b}}{\hspace{5pt}} &
\multicolumn{1}{c}{Filter}&
\multicolumn{1}{c}{Exposure}&
\multicolumn{1}{c}{Date}\\
\multicolumn{1}{c}{} &
\multicolumn{2}{c}{(J2000.0)} &
\multicolumn{1}{c}{} &
\multicolumn{1}{c}{} &
\multicolumn{1}{c}{}& 
\multicolumn{1}{c}{$\sec$}&
\multicolumn{1}{c}{Observ.}\\
\hline 
RCW\,87            &15 05 20.0&$-$57 31 15&320.15&   0.79&$J,K_S$  &300&25.06.2004\\  
$[$DBSB2003$]$\,172&16 41 10.0&$-$47 07 27&337.92&$-$0.47&$H,K_S$  &300&25.06.2004\\ 
$[$BDSB2003$]$\,164&17 25 32.0&$-$36 21 58&351.47&$-$0.46&$J,H,K_S$&300&25.06.2004\\  
\hline
\end{tabular}
\end{center}
\end{table*}

The stellar photometry of the final images was carried out with 
ALLSTAR in DAOPHOT\,II (Stetson \cite{ste93}). We considered 
only stars with DAOPHOT errors less than 0.2\,mag. The median 
averaged internal photometric errors are $0.03\pm0.02$ for the 
$J$, $H$, $K_S$ stars brighter than 17\,mag and $0.07\pm0.04$ 
for the fainter ones. We added to the errors in quadrature an 
extra observational uncertainty of $\sim$0.03\,mag due to the 
sky background variations. We replaced the brightest 
stars (usually with $K_S$$<$12\,mag), that were saturated on 
our images, with the 2\,MASS measurements.

The weather conditions were nonphotometric (typical seeing 
1-1.2\,arcsec) during all our observing runs, forcing us to 
calibrate the data by comparing our instrumental magnitudes 
with the 2\,MASS magnitudes of 10-25 stars per image, depending 
on the field crowding and the band. The standard error values 
for the coefficients are less than 0.03 for the zero-point and 
less than 0.02 for the colour term. Our conservative
estimate of the total external errors of our photometry is 
0.04-0.05\,mag.

\section{[BDB2003]\,0320.15+00.79 or RCW\,87}

The IR star cluster candidate RCW\,87 was selected from the Bica et 
al. (2003a) catalog. It is 
surrounded by CH87 320.153+0.780 (RCW\,87) H\,{\sc II} region 
(Rodgers et al. \cite{rod60}). Caswell \& Haynes (\cite{cas87}) 
measured a radial velocity V$_r$=$-$36.0$\pm$1\,km/s from the 
H$_2$CO and based on a Galaxy rotation model they determine a 
distance between 2.5 and 12.9\,kpc, giving preference to the 
smaller value. Later, Simpson \& Rubin (\cite{sim90}) measured 
the luminosity of the H\,{\sc II} region $L/L_{\odot}$=2.4$ \times$10$^4$ from
an improved Galactic rotation model.

\subsection{Near-IR $JK_S$ imaging}
We covered RCW\,87 with two overlapping pointings. The $K_S$ 
images are shown on Fig.~\ref{fig01} and Fig.~\ref{fig02}. A 
pseudo-true colour image is shown in Fig.~\ref{fig03}.

\begin{figure}
\caption{The $K_S$-band image of RCW\,87 - Field\,1. 
  The field of view is 2.1$\times$2.1 arcmin. North is up, 
  and East is to the left.}
\label{fig01}
\end{figure}
\begin{figure}
\caption{The $K_S$-band image of RCW\,87 - Field\,2. 
  See Fig.~\ref{fig01} for details.}
\label{fig02}
\end{figure}

\begin{figure}
\caption{The pseudo-true colour image of RCW\,87. 
  The field of view is $\sim$3.8$\times$2.1 arcmin. 
  North is up, and East is to the left.
  Blue corresponds to $J$-band, green is the average of 
  the $J$ and $K_S$ images, and red is the $K_S$ band.}
\label{fig03}
\end{figure}

\begin{figure}
\resizebox{\hsize}{!}{\includegraphics{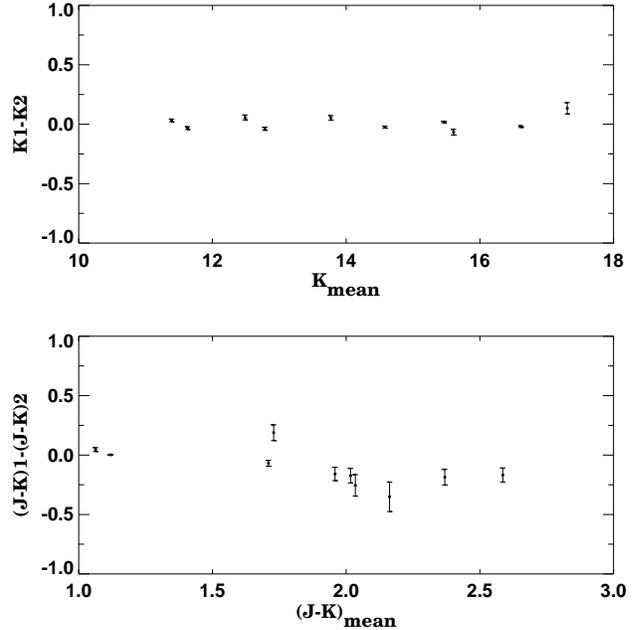}}
\caption{The mean value versus difference in the measurements 
of the $K_s$ magnitude and $J-K_s$ colours of the common stars between 
two observed fields of RCW\,87.
}
\label{fig04}
\end{figure}

We used the common stars on the two pointings to estimate the 
errors of our photometry. The difference versus the mean value 
plots for the $K_S$-band and the $J$$-$$K_S$ colour are shown in 
Fig.~\ref{fig04}. The r.m.s. for the $K_S$-band is 0.02\,mag,
and 0.06\,mag for the colour, with a small systematic trend. 
Thus, we transformed the ($J$$-$$K_S$) of the first field to the 
second one with the equation: 
($J$$-$$K_S$)$_{\rm Field\,2}$=$-$0.214+1.17*($J$$-$$K_S$)$_{\rm Field\,1}$.
We attribute the difference to the changing weather conditions
during the observations.

Our final photometry list contains 1471 stars. The three brightest stars 
in the field are saturated on our images and their
 magnitudes were taken from the 2\,MASS Point 
Source Catalog. The $K_S$ versus 
$J$$-$$K_S$ colour-magnitude diagram (CMD) of all stars is shown 
in Fig.~\ref{fig05} (left).
 We also observed a comparison field 
for back-/foreground subtraction with the same integration time
and field of view,
at RA=15:05:14.3 and DEC=$-$57:28:8.7 (J2000). The CMD of that 
field is also shown on Fig.~\ref{fig05} (right). As can be seen, 
the field stars have $J$$-$$K_S$$<$1.5\,mag, while the cluster 
stars occupy the region 1.5$<$$J$$-$$K_S$$<$3.0\,mag. There are 
also stars with strong IR excess at $J$$-$$K_S$$>$3.0\,mag. One 
of the three brightest stars at RA=15:05:20.566 and 
DEC=$-$57:30:57.5 (J2000) has $J$$-$$K_S$=1.08\,mag and 
$K_S$=8.57\,mag and it is obviously a field star, while the 
others at RA=15:05:25.04 and DEC=$-$57:30:55.96, and 
RA=15:05:16.9, DEC=$-$57:30:05.67 (J2000) have 
$J$$-$$K_S$=6.06\,mag, $K_S$=7.89\,mag and 
$J$$-$$K_S$=7.73\,mag, $K_S$=8.87\,mag, respectively, and are the 
reddest stars in the CMD. 
We statistically decontaminated the cluster CMD from the field 
population using the comparison field, leaving only 1121 probable 
cluster members. The details of the procedure are given in 
Borissova et al. (2005).

\begin{figure}
\caption{Left panel: The $K_S$ versus $J$$-$$K_S$ CMD of 
\object{RCW\,87}. All stars in our photometry list are shown 
with solid dots. The unreddened Main Sequence (Schmidt-Kaler 
\cite{sch82}) is drawn with a solid line, and with dashed 
line for reddening corresponding to E($B$$-$$V$)=3.4\,mag or
A$_V$=10.9\,mag. The reddening vectors for O5\,V and B0\,V 
stars are also shown. A distance modulus of 
($m$$-$$M$)$_0$=17.42\,mag was adopted (see the text).
Right panel: The $K_S$ versus $J$$-$$K_S$ CMD for the 
comparison field.
}
\label{fig05}
\end{figure}

\begin{figure}[htbp]
\resizebox{\hsize}{!}{\includegraphics{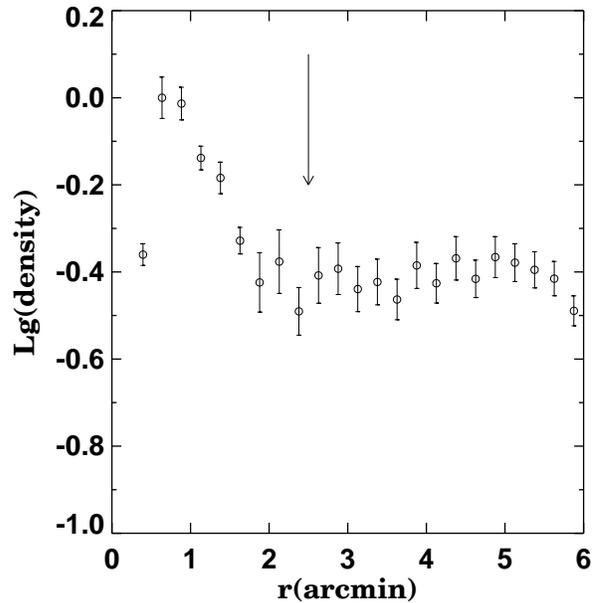}}
\caption{Radial profile of RCW\,87. The projected $K_s$ 
star number density (number of stars per square arcmin) is 
plotted versus the radius fro the cluster center. The bars 
show the $3\sigma$ Poisson uncertainties. The arrow 
indicates the determined 2.5\,arcmin cluster radius.}
\label{fig06}
\end{figure}

To determine the boundaries of the cluster we performed direct 
star counting, assuming spherical symmetry. Since the field of 
view of Panic covers the object only partially, we used the 
2\,MASS photometry within 10\,arcmin from the cluster center. The 
cluster boundary was defined as the point of the profile where 
the excess density becomes twice the standard deviation of the 
surface density of the surrounding field. This yields a radius 
of 2.5\,arcmin. Fig.~\ref{fig06} shows the projected $K_s$ 
band star number density (number of stars per sq. arcmin) as a
function of the radius. The depression near the cluster center 
is due to crowding in the 2\,MASS.

\subsection{Near-IR Spectroscopy}

Near-IR spectra in the K atmospheric window of three stars 
(marked with squares in Fig.~\ref{fig05}) in the field of 
RCW\,87 were obtained with SofI at the ESO NTT. We used a 
grism yielding a resolution R$\sim$2200 over 
$\lambda\lambda$\,2.10-2.35$\mu$m range, and 1\,arcsec wide 
slit. The observations were made with nodding between two 
positions along the slit. The poor weather conditions made 
it possible to use only three images of 120\,sec integration 
each. The data were reduced in the typical manner: sky 
subtraction, extraction of 1-dimensional spectra, wavelength 
calibration, division by a telluric standard (HIP\,78169, 
G1\,V) and removal of the artificial emission lines created 
by the telluric (Maiolino et al. \cite{mai96}). 

\begin{figure}[htbp]
\resizebox{\hsize}{!}{\includegraphics{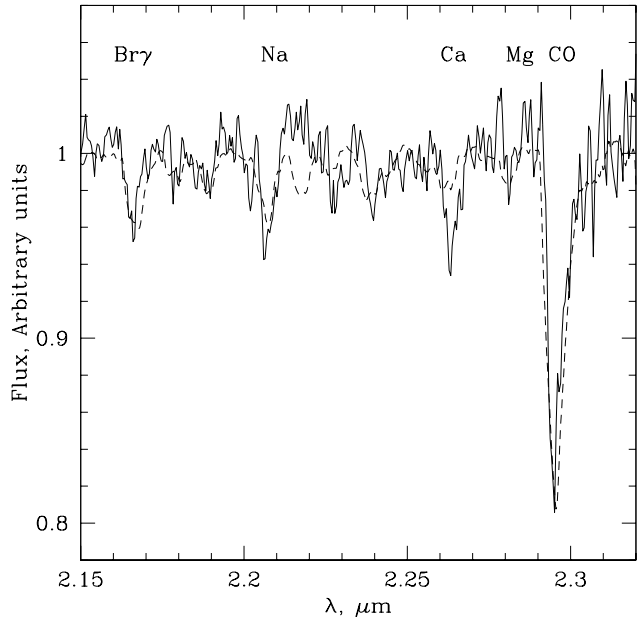}}
\caption{The K spectrum of the star at RA=15:05:22, 
DEC=$-$57:31:27 (J2000) star in RCW\,87. The reference 
spectrum for K0.5\,II star from Ivanov et al. 
(\cite{iva04}) is overplotted with a dashed line.
}
\label{fig07}
\end{figure}

The spectrum of the brightest star (RA=15:05:22, 
DEC=$-$57:31:27, J2000) is plotted on Fig.~\ref{fig07}. 
Since there is no clear scheme for spectral classification 
of red stars in the near-IR we simply cross-correlated the
spectrum with the red stars in the spectral library of 
Ivanov et al. (\cite{iva04}). The best fit was obtained for 
a K0.5\,II star. Note that metallicity effects are not taken 
into account, and the comparison with other spectra 
indicated that the error of the spectral type determination 
is three  sub-types.

The comparison between the observed and the intrinsic 
colours, and the apparent and absolute magnitudes for the K0 
supergiant -- ($J$$-$$K$)$_0$=0.58\,mag and $M_k$=$-$4.66\,mag 
from Koornneef (\cite{kor83}) -- yields: reddening of 
E($J$$-$$K_S$)=1.98\,mag or E($B$$-$$V$)=3.41\,mag and 
distance modulus of ($m$$-$$M$)$_0$=17.42\,mag (D=7.6\,Kpc).
These values were used to plot the unreddened main sequence 
from Schmidt-Kaler (\cite{sch82}) on the CMD shown in 
Fig.~\ref{fig05}. As can be seen, the main sequence stars 
form a well-defined locus and the brightest cluster members 
are evolved red supergiants.

The two other spectra of fainter stars that fell in the 
slit serendipitously were disregarded because they had 
insufficient signal-to-noise for spectral classification.

We estimate the age of the cluster by comparison with isochrones
from the Padova stellar library (Bonatto et al. 2004), which 
provided theoretical isochrones computed for the 2\,MASS 
photometric system. We used the distance and the reddening listed 
above and solar metallicity. The best fit favors a cluster age 
between 20 and 25 Myr (Fig.~\ref{fig08}). 

\begin{figure}
\vspace{4mm}
\caption{The decontaminated $M_K$, ($J$$-$$K_S$)$_0$ colour 
magnitude diagram of RCW\,87 with superimposed isochrones from 
the Padua library (Bonatto et al. 2004).}
\label{fig08}
\end{figure}

The total cluster mass was calculated as described in Borissova 
et al (2003): we adopted the 20 Myr isochrone, and counted 
cluster stars between reddening lines originating from positions 
on the isochrone for different initial masses. The mass interval 
is relatively narrow: from 4.5 to 11 solar masses, so we refrain
from determining the initial mass function slope. The minimal 
total cluster mass, equal to the sum of the masses of suspected 
cluster members is $\sim$10300 solar masses. 

Given the nature of our distance estimate we cannot obtain a 
formal error estimate but taking into account the difference in
the luminosity of stars with the nearest spectral classes, we 
tentatively assume that our distance has 20\% uncertainty. This
corresponds to about 300 $M_{\odot}$ variation in the total 
cluster mass.

\subsection{Mid-IR imaging}

We used the 3.6, 4.5, 5.8 and 8.0\,$\mu$m observations of 
RCW\,87 obtained with the Infrared-Array Camera on the {\it 
Spitzer Space Telescope} as a part of the Galactic Legacy 
Infrared Mid-Plane Survey Extraordinaire (hereafter GLIMPSE; 
Benjamin et al. 2003) to search for pre-main sequence 
cluster members. Wilking \& Lada (1983) developed a 
classification scheme of young stellar objects based on the 
slope of their mid-IR Spectral Energy Distributions (SEDs). 
Later, it was placed in an evolutionary context by Adams et 
al. (\cite{ada87}) who modeled the SEDs as stars surrounded 
by dusty disks and envelopes. The comparison of these models 
with the observed SEDs suggested that the Class\,I objects 
are protostars with infalling envelopes, and Class\,II 
objects are stars with disks. Class\,III objects have the 
SEDs of stellar photospheres. The three classes form an 
evolutionary sequence, with young stars evolving from 
Class\,I to Class\,II and finally to Class\,III objects 
(Kenyon \& Hartmann \cite{ken95}). 

The cluster is clearly visible on [3.6] and [4.5] $\mu$m, 
while the other bands are dominated by dust emission (see 
Fig.~\ref{fig09} and Fig.~\ref{fig10}).
In total, 171 stars within radius 2.5\,arcmin around the 
cluster center have at least one mid-IR measurement (marked 
with squares on Fig.~\ref{fig09}). The [4.5], [3.6-4.5]
colour magnitude diagram and [5.8-8.0], [3.6-4.5] colour-colour 
diagram are shown in Fig.~\ref{fig11}.
Most RCW\,87 stars that cluster around [3.6]$-$[4.5]=0\,mag 
and [5.8]$-$[8.0]=0\,mag appear to be field stars or 
Class\,III sources, with no intrinsic IR excess. We 
identified three Class\,II objects and two Class\,I 
objects.

\begin{figure}
\caption{The [3.6]-band image of RCW\,87. 
  The field of view is 5.0$\times$5.0 arcmin. North is up, 
  and East is to the left. Probable cluster members within 
  radius 2.5\,arcmin are marked with squares.}
\label{fig09}
\end{figure}
\begin{figure}
\caption{The [[3.6],[4.5] and [5.8] (shown in blue, green and red,
  respectively) true colour image of RCW\,87. The field of view is
  5.0$\times$5.0\,arcmin. North is up, and East is to the left.}
\label{fig10}
\end{figure}
\begin{figure}
\resizebox{\hsize}{!}{\includegraphics{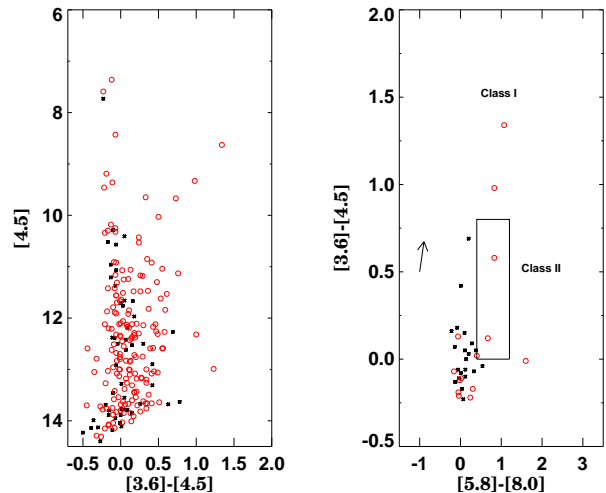}}
\caption{Mid-IR colour-magnitude and colour-colour diagrams for 
RCW\,87. Probable cluster members (stars within 2.5\,arcmin from
the cluster center) are plotted with open red circles,
comparison field stars are shown as solid asterisks. The arrows 
are the reddening vectors for A$_V$=11\,mag (Fitzpatrick 
\cite{fit99}). The box marks the location of Class\,II objects 
(Allen et al. \cite{all04}). Class\,I objects are above and to 
the right of the square.}
\label{fig11}
\end{figure}

\section{[BDSB2003]\,164}

The IR star cluster candidate [BDSB2003]\,164 (Fig.~\ref{fig12})
was selected from Bica et al. (\cite{bic03b}). It is located 
in a crowded field, and it is surrounded by an H\,{\sc II} 
region IRAS\,17221-3619. Peeters et al. (\cite{pet02}) measured a 
radial velocity of $-$21.0\,km\,s$^{-1}$ from IR recombination 
lines. They derived a distance of 3.4\,Kpc and a luminosity of
$L/L_{\odot}$=7.52$\times$10$^4$.

\begin{figure}
\vspace{0.5cm}
\caption{The true colour image of [BDSB2003]\,164. 
  The field of view is 2.1$\times$2.1 arcmin. North is up, 
  and East is to the left.}
\label{fig12}
\end{figure}

\begin{figure}
\resizebox{\hsize}{!}{\includegraphics{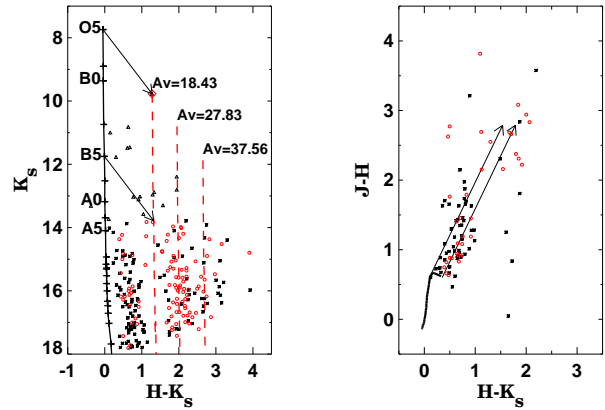}}
\caption{Left panel: The $K_S$ versus $H$$-$$K_S$ 
colour-magnitude diagram of [BDSB2003]\,164.
Right panel: The $H$$-$$K_S$ versus $J$$-$$H$ colour-colour diagram.
Probable cluster members are plotted with open red circles, while
comparison field stars are shown as black asterisks. The triangles
represent stars from 2\,MASS catalog. The blue square show the star
for which we have spectra.
The unreddened Main Sequence (Schmidt-Kaler \cite{sch82}) is 
shown with a solid line, arrows are reddening vectors for 
A$_V$=18.43\,mag.
With a dashed line are shown Main Sequences for different reddening 
values, corresponding to E($B$$-$$V$)=5.76, E($B$$-$$V$)=8.70, and 
E($B$$-$$V$)=11.74\,mag. 
}
\label{fig13}
\end{figure}

The object is practically invisible on the $J$-band images, but 
the overdensity on the $H$ and $K_s$-bands is obvious. Based on 
the $K_s$ profile we adopted a radius of 0.3 arcmin. The $K_S$ 
versus $H$$-$$K_S$ colour-magnitude diagram of all stars in our 
field (2785 stars) is shown in Fig.~\ref{fig13}, left panel. To 
obtain an estimate of the fore- and background contamination we 
define a non-cluster region, with radius greater than 0.7 arcmin 
and an area equal to that of the cluster. The stars within the 
adopted cluster radius are shown with open red circles, while 
the comparison field stars are plotted as black asterisks. The 
brightest stars are saturated and their magnitudes are taken 
from 2\,MASS (shown in Fig.~\ref{fig13} as triangles). Most 
of the field stars in the CMD have $H$$-$$K_S$$<$1.0\,mag, 
while the stars within a 0.3 arcmin radius are located between 
1.5$<$$H$$-$$K_S$$<$3.0\,mag. The ($H$$-$$K_s$) vs. ($J$$-$$H$) 
colour-colour diagram is plotted on the right panel. Two reddening 
vectors for $A_v=20$, encompassing the main sequence locus, are 
drawn as arrows on the colour-colour diagram. The 
unreddened Main Sequence (Schmidt-Kaler \cite{sch82}) is shown as 
a solid line. We used the absorption ratios from Bessell et al. 
(1998). The colour-colour diagram reveals the severe and 
variable extinction -- almost half of the stars are reddened 
main-sequence stars. Some stars with strong IR excess and 
probably pre-main sequence stars. 
After statistically cleaning the cluster CMD by removing from the 
cluster colour-magnitude diagram as many stars as there are present 
on the ``field'' colour-magnitude diagram, we selected 74 probable 
cluster members. 

\begin{figure}[htbp]
\resizebox{\hsize}{!}{\includegraphics{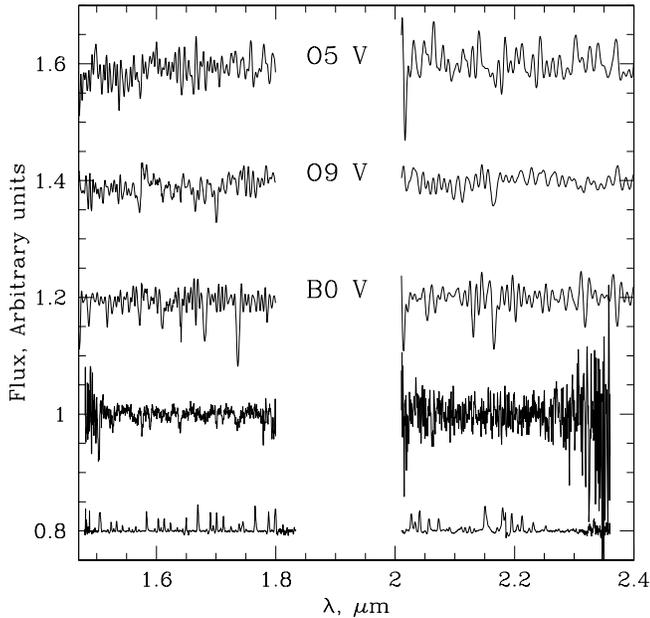}}
\caption{The IR spectrum of the RA=17:25:32 and DEC=$-$36:21:50 star.
The H- (left) and K-band (right) spectra of O5\,V, O9\,V and B0\,V 
from the Pickles (1998) library are shown in the first three positions, 
followed by the program star and spectrum of the sky.
}
\label{fig14}
\end{figure}

The reduced NTT spectrum of the star with coordinates: RA=17:25:32 
and DEC=$-$36:21:50 is plotted in Fig.~\ref{fig14}. The star is 
marked with a square on the colour magnitude diagram and lies 
away from the locus of the main sequence stars. In the true 
colour image, however we can see that it is in the middle of the 
cluster and probably is its brightest cluster member. To obtain 
the spectral type we compared our spectrum with the stellar 
library of Pickles (1998). The H- (Fig.~\ref{fig14}, left) and 
K-band (Fig.~\ref{fig14}, right) spectra of O5\,V, O9\,V and 
B0\,V from this library are shown in the first three positions, 
followed by the program star and spectrum of the sky. Most likely 
the target star belongs to the O5\,V -- 09\,V sub-class. The intrinsic 
IR colour and the absolute magnitudes for O5\,V and O9\,V stars 
are: ($H$$-$$K_S$)$_0$=$-$0.05, $M_k$=$-$4.79\,mag, and 
($H$$-$$K_S$)$_0$=$-$0.04, $M_k$=$-$3.63\,mag (Schmidt-Kaler 
\cite{sch82}), respectively. If the stellar type is O9\,V, the 
calculated reddening and distance are E($B$$-$$V$)=5.72 and 
($m$$-$$M$)$_0$=11.32\,mag (D=1.9 Kpc), much smaller than the 
distance of D=3.4\,Kpc to the H{\sc ii} region (Peeters et al. 
\cite{pet02}). Assuming O5\,V type leads to E($B$$-$$V$)=5.76 and 
($m$$-$$M$)$_0$=12.53\,mag (D=3.2 Kpc), fully comparable with the 
those for the H\,{\sc II} region. However, plotting 
in Fig.~\ref{fig13} the reddened main sequence of Schmidt-Kaler 
(\cite{sch82}) shows that the low luminosity cluster members are
redder than expected. Most likely this is due to differential 
reddening -- the strong wind of the O star has cleared the dust 
from its immediate surroundings while low mass stars are still 
dust-embedded. Therefore, we adopted for the rest of the cluster 
E($B$$-$$V$)=8.70\,mag.

Adopting the 1\,Myr isochrone we derived 
a minimal total cluster mass, comprised 
of the mass of suspected cluster members of 70 solar masses. 
An integration of the extrapolated power-law IMF fit down to 
0.08 $M_{\odot}$ yields a total mass of 1760 $M_{\odot}$ which 
should be considered as an upper limit.

\section{[DBSB2003]\,172}

The IR star cluster candidate [DBSB2003]\,172 was selected from 
Dutra et al. (\cite{dut03}). It is surrounded by the 
H\,{\sc II} region GAL337.9-00.5 with a kinematical distance from 
the Sun of 3.0\,Kpc and a galactocentric radius of 5.8\,Kpc 
(Giveon et al. 2002).

The $K_S$ band image of our fields is shown in Fig.~\ref{fig15} 
and the  $K_S$ versus $H$$-$$K_S$ colour-magnitude diagram of all 
3564 stars in our photometry list is shown in Fig.~\ref{fig16}. 
The unreddened Main Sequence (Schmidt-Kaler \cite{sch82}) is 
plotted with a solid line, and with dashed lines for different 
reddening values, corresponding to E($B$$-$$V$)=2.0 and 
E($B$$-$$V$)=4\,mag. The reddening vectors for B0\,V and B5\,V 
stars are also shown, for A$_V$=17.6\,mag. The 
$(m-M)_0$=12.39\,mag is adopted from Giveon et al. (2002). 

There is no overdensity of stars on our 300 sec images. The 
colour-magnitude diagram shows fore/background stars visible 
around $H$$-$$K_S$=0.5\,mag and reddened Main Sequence stars 
with $A_V$=6.4\,mag in the region between $H$$-$$K_S$=1.5 and 
2.5\,mag. There are 23 stars with $H$$-$$K_S$$>$2.5\,mag. They 
are overplotted on Fig.~\ref{fig15} and closely follow  the 
morphology of the gas.

Thus, [DBSB2003]\,172 shows only a weak concentration of red 
stars, associated with the H\,{\sc II} region and most likely 
this is an extended star forming region. 
 
\begin{figure}
\caption{The $K_S$-band image of [DBSB2003]\,172. 
  The field of view is 2.1$\times$2.1 arcmin. North is up, 
  and East is to the left.}
\label{fig15}
\end{figure}

\begin{figure}
\caption{The $K_S$ versus $H$$-$$K_S$ colour-magnitude diagram 
of [DBSB2003]\,172. The unreddened Main Sequence (Schmidt-Kaler 
\cite{sch82}) is shown with a solid line, and with dashed lines 
for different reddening values, corresponding to E($B-V$)=2.0, 
and E($B-V$)=4\,mag. 
The reddening vectors for B0\,V and B5\,V stars are also shown, for 
A$_V$=17.6\,mag. 
The $(m-M)_0$=12.39\,mag is adopted from Giveon et al. (2002).
}
\label{fig16}
\end{figure}

Mid-IR 3.6, 4.5, 5.8 and 8.0\,$\mu$m imaging of [DBSB2003]\,172 is 
available from the Spitzer Space Telescope. A [3.6], [5.8] and 
[8.0] $\mu$m true colour image is shown on Fig.~\ref{fig17}. Again, 
there is no obvious centrally concentrated overdensity of the stars.
 
\begin{figure}
\caption{The [[3.6], [4.5] and [5.8] true colour image of 
  [DBSB2003]\,172. 
  The field of view is $\sim$5.0$\times$5.0 arcmin. North is up, 
  and East is to the left.}
\label{fig17}
\end{figure}

The mid-IR photometry of stars within a 5 arcmin radius around 
[DBSB2003]\,172 is shown in Fig.~\ref{fig18}. The [4.5] vs [3.6]$-$[4.5] 
colour-magnitude diagram is shown in the top left panel, the 
[4.5]$-$[5.8] versus [3.6]$-$[4.5],[5.8]$-$[8.0] versus [4.5]$-$[5.8] 
and [5.8]$-$[8.0] versus [3.6]$-$[4.5] colour-colours diagrams are shown 
in the other panels. Most of the stars cluster around [3.6]$-$[4.5]=0 
and [5.8]$-$[8.0]=0 and are background/foreground stars or Class\,III 
sources with no intrinsic IR excess. Fifteen stars have 
0.0$<$[3.6]$-$[4.5]$<$0.8 and 0.4$<$[5.8]$-$[8.0]$<$1.1 and therefore 
they are\,Class II pre-main sequence objects (Allen et al. \cite{all04}).  
There are also three objects, that  can be attributed to Class\,I. All sources
mentioned are located outside the field of view of our near-IR 
observations.

\begin{figure}
\caption{[DBSB2003]\,172 in the mid-IR.
Top left: The [4.5] versus [3.6]$-$[4.5] colour-magnitude diagram. 
Top right: The [4.5]$-$[5.8] versus [3.6]$-$[4.5] colour-colour diagram.
Bottom left and bottom right panels show [5.8]$-$[8.0] versus 
[4.5]$-$[5.8] and [5.8]$-$[8.0] vs [3.6]$-$[4.5] colour-colours 
diagrams. The arrows indicate A$_V$=30\,mag reddening vectors 
(Fitzpatrick 1999). The box marks the location of Class\,II objects 
as defined in Allen et al. (2004).}
\label{fig18}
\end{figure}

\section{Summary}

We report photometric and spectroscopic observations of three 
embedded cluster candidates located in the general direction of
the Galactic Center. Two of them -- RCW\,87 and [BDSB2003]\,164 -- 
are genuine clusters, while [DBS2003]\,172 is probably an extended 
star forming region.

Our deep $J$ and $K_S$ images indicate that RCW\,87 is a dense 
stellar cluster with a radius of $\sim$2.5 arcmin. The K-band 
spectrum of the brightest cluster member identifies it as a 
K0.5\,II supergiant, yelding E($B$$-$$V$)=3.4, A$_V$=10.9, and 
$(m-M)_0$=17.42\,mag (D=7.6 kpc). The cluster age is between 20 
and 25\,Myr and its brightest members have already evolved into 
red supergiants. The isochrone analysis indicated 
that the cluster is massive -- the total mass of the detected 
cluster members is 10300 solar masses, in the range of the
most massive known young clusters in the Milky Way. Further analysis 
of this object is necessary to confirm this
interesting result. The mid-IR photometry from the 
{\it Spitzer Space Telescope} shows the presence of at least five 
Class\,I and II proto-stellar objects indicating that some 
triggered star formation in the  H\,{\sc II} region is still going
on in the vicinity of this cluster.

The cluster candidate [BDSB2003]\,164 is a small star cluster 
with only 74 probable members within a 0.3 arcmin radius. 
The brightest cluster star is identified as an O5 main sequence 
star, based on our IR spectrum. We calculated a distance modulus 
of $(m-M)_0$=12.53\,mag (R=3.2 kpc) and reddening of 
E($B$$-$$V$)=8.7\,mag. The mass of the detected cluster members is 
$\sim$70 solar masses and an upper mass limit of 1760 solar masses
is estimated.

The deep $J$, $H$ and $K_S$ images show no concentration of stars 
outlining a cluster for the [DBSB2003]\,172 cluster candidate. 
The stars with the reddest colours follow 
closely the morphology of the gas. On the other hand, the mid-IR 
photometry shows active on-going star formation in the H\,{\sc II} 
region.

\begin{acknowledgements}
This publication makes use of data products from the Two Micron 
All Sky Survey, which is a joint project of the University of 
Massachusetts and the Infrared Processing and Analysis 
Center/California Institute of Technology, funded by the 
National Aeronautics and Space Administration and the National 
Science Foundation. This research has made use of the SIMBAD 
database, operated at CDS, Strasbourg, France. This work is 
based [in part] on observations made with the {\it Spitzer Space 
Telescope}, which is operated by the Jet Propulsion Laboratory, 
California Institute of Technology under a contract with NASA."
The authors gratefully acknowledge the comments 
by the anonymous referee. D.M. and D.G. are supported by 
FONDAP Center for Astrophysics grant number 15010003. 
\end{acknowledgements}


\begin{thebibliography}{}
\bibitem[2004]{all04} Allen, L., Calvet, N., D'Alessio, P., 
  Hartmann, L., Megeath, S. et al. 2004, AlS, 154, 363
\bibitem[1987]{ada87} Adams, F.C., Lada, C.J., \& Shu, F.H. 
  1987, ApJ, 312, 788
\bibitem[2003]{ben03} Benjamin, R., Churchwell, E., Babler, 
  L., Bania, T. M., Clemens, P. et al. 2003, PASP, 115, 953
\bibitem[2003a]{bic03a} Bica, E., Dutra, C.M. \& Barbuy, B., 
  2003a, \aap, 397, 117
\bibitem[2003b]{bic03b} Bica, E., Dutra, C.M., Soares,J. \& 
  Barbuy, B. 2003b, \aap, 404, 223
\bibitem[1998]{bes98} Bessell, M.S., Castelli, F. \& Plez, B. 
  1998, \aap, 333, 231
\bibitem[2004]{} Bonatto, Ch., Bica, e., Girardi, L., 2004,  \aap, 415, 571 
\bibitem[2003]{bor03} Borissova, J., Pessev, P., Ivanov, V.D., 
  Saviane, I., Kurtev, R., Ivanov, G.R. 2003, \aap, 411, 83 
\bibitem[2003]{bor05} Borissova, J., Ivanov, V.D., Minniti, 
  D., Geisler, D., Stephens, A. W. 2005, \aap, 435, 95 
\bibitem[1987]{cas87} Caswell, J.L. \& Haynes, R.F. 1987, 
  A\&A, 171, 261
\bibitem[1996]{chan96} Chan, S., Henning, T., Schreyer, K, 
  1996, A\&AS, 115, 285   
\bibitem[2002]{giv02} Giveon U., Sternberg A., Lutz D., 
  Feuchtgruber H., Pauldrach A.W.A., 2002, AJ, 566, 880 
\bibitem[1999]{fit99} Fitzpatrick, E., 1999, PASP, 111, 63
\bibitem[2003]{dut03} Dutra, C.M., Bica, E., Soares, J. \& 
  Barbuy, B. 2003, \aap, 400, 533
\bibitem[1997]{epc97} Epchtein, N. 1997, in ASSL Vol. 210, 
  The Impact of Large Scale Near-IR Sky Surveys, ed. F. Garzon 
  et al. (Dordrecht: Kluwer), 15
\bibitem[2002]{iva02} Ivanov, V.D., Borissova, J., Pessev, P., 
  Ivanov, G.R., Kurtev, R. 2002, \aap, 394, 1  
\bibitem[2002]{iva05} Ivanov, V. D., Borissova, J., Bresolin, F., 
  Pessev, P. 2005, \aap, 435, 107  
\bibitem[2004]{iva04} Ivanov, V. D., Rieke, M.J, Engelbracht, C.W. 
  Alonso-Herrero, A., Rieke, G.H. \& Luhman, K.L. 2004, \apj, 151, 
  387
\bibitem[1995]{ken95} Kenyon, S. J., \& Hartmann, L. 1995, ApJS, 
  101, 117
\bibitem[1983]{kor83} Koornneef, J.,  1983, \aap, 128, 84
\bibitem[1997]{kuc97} Kuchar, T.A.,Clark, F.O., 1997, ApJ, 488, 224
\bibitem[1996]{mai96} Maiolino, R., Rieke, G.H. \& Rieke, M.J. 1996,
  \aj, 111, 537
\bibitem[2002]{pet02} E. Peeters, E., Martin-Hernandez, N, 
  Damour, F., Cox, E., Roelfsema,P.R. et al., 2002, A\&A 381, 
  571  
\bibitem[1998]{} Pickles, A.J., 1998, PASP, 110, 863
\bibitem[1960]{rod60} Rodgers, A.W., Campbell, C.T., Whiteoak, 
  J.B., 1960, MNRAS, 121,103
\bibitem[1990]{sim90} Simpson, J.P., Rubin, R.H., 1990, AJ, 354, 165
\bibitem[1982]{sch82} Schmidt-Kaler, T., 1982, in Landolt-Borstein, 
  New Series, Group VI, vol. 2, ed. K. Schaifers \& H.H. Voigt 
  (Berlin: Springer-Verlag),1 
\bibitem[1997]{scr97} Skrutskie, M.F., et al. 1997, in ASSL Vol. 
  210, The Impact of Large Scale Near-IR Sky Surveys, ed. F. 
  Garzon et al. (Dordrecht: Kluwer), 25
\bibitem[1993]{ste93} Stetson, P. B. 1993, User's Manual for 
  {\sc daophot ii}
\bibitem[1983]{} Wilking, B. A., \& Lada, C. J. 1983, ApJ, 274, 698
\end{thebibliography}
\end{document}